\documentclass[aps,prappl,twocolumn,groupedaddress,superscriptaddress,floatfix,longbibliography]{revtex4-1}

\usepackage{epsfig}
\usepackage{multirow}
\usepackage{amsmath,amssymb,mathtools}
\usepackage{color}
\usepackage{graphicx}
\usepackage{dcolumn}
\usepackage{bm}
\usepackage{subfigure}
\usepackage[utf8]{inputenc}
\usepackage[T1]{fontenc}
\usepackage{tikz}

\begin{document}

\title{rf-SQUID measurements of anomalous Josephson effect}

\author{C. Guarcello}
\affiliation{Dipartimento di Fisica ``E.R. Caianiello'', Universit\`a di Salerno, Via Giovanni Paolo II, 132, I-84084 Fisciano (SA), Italy}
\affiliation{Centro de Física de Materiales, Centro Mixto CSIC-UPV/EHU, Paseo Manuel de Lardizabal 5, 20018 San Sebastián, Spain}
\author{R. Citro}
\affiliation{Dipartimento di Fisica ``E.R. Caianiello'', Universit\`a di Salerno, Via Giovanni Paolo II, 132, I-84084 Fisciano (SA), Italy}
\affiliation{Spin-CNR, Universit\`a di Salerno, I-84084 Fisciano (SA), Italy}
\affiliation{INFN, Sezione di Napoli Gruppo Collegato di Salerno, Complesso Universitario di Monte S. Angelo, I-80126 Napoli, Italy}
\author{O. Durante}
\affiliation{Dipartimento di Fisica ``E.R. Caianiello'', Universit\`a di Salerno, Via Giovanni Paolo II, 132, I-84084 Fisciano (SA), Italy}
\affiliation{INFN, Sezione di Napoli Gruppo Collegato di Salerno, Complesso Universitario di Monte S. Angelo, I-80126 Napoli, Italy}
\author{F.S. Bergeret}
\affiliation{Centro de Física de Materiales, Centro Mixto CSIC-UPV/EHU, Paseo Manuel de Lardizabal 5, 20018 San Sebastián, Spain}
\affiliation{Donostia International Physics Center, Paseo Manuel de Lardizabal 4, 20018 San Sebastián, Spain}
\author{A. Iorio}
\affiliation{NEST, Istituto Nanoscienze-CNR and Scuola Normale Superiore, Piazza San Silvestro 12, I-56127 Pisa, Italy}
\author{C. Sanz-Fern\'andez}
\affiliation{Centro de Física de Materiales, Centro Mixto CSIC-UPV/EHU, Paseo Manuel de Lardizabal 5, 20018 San Sebastián, Spain}
\author{E. Strambini}
\affiliation{NEST, Istituto Nanoscienze-CNR and Scuola Normale Superiore, Piazza San Silvestro 12, I-56127 Pisa, Italy}
\author{F. Giazotto}
\affiliation{NEST, Istituto Nanoscienze-CNR and Scuola Normale Superiore, Piazza San Silvestro 12, I-56127 Pisa, Italy}
\author{A. Braggio}
\affiliation{NEST, Istituto Nanoscienze-CNR and Scuola Normale Superiore, Piazza San Silvestro 12, I-56127 Pisa, Italy}


\begin{abstract}
We discuss the response of an rf-SQUID formed by anomalous Josephson junctions embedded in a superconducting ring with a non-negligible inductance. We demonstrate that a properly sweeping in-plane magnetic field can cause both the total flux and the current circulating in the device to modulate and to behave hysteretically. The bistable response of the system is analyzed as a function of the anomalous phase shift at different values of the screening parameter, in order to highlight the parameter range within which a hysteretic behavior can be observed. The magnetic flux piercing the SQUID ring is demonstrated to further modulate the hysteretical response of the system. Moreover, we show that the anomalous phase shift can be conveniently determined through the measurement of the out-of-plane magnetic field at which the device switches to the voltage state and the number of trapped flux quanta changes. Finally, we compare the response of two different device configurations, namely, a SQUID including only one or two anomalous junctions. In view of these results, the proposed device can be effectively used to detect and measure the anomalous Josephson effect. 
\end{abstract}

\maketitle

\section{Introduction}
\label{section0}\vskip-0.2cm

A superconducting quantum-interference device (SQUID), which is formed embedding Josephson junctions (JJs) in a phase-sensitive superconducting loop geometry, is an efficient and versatile tool to measure phase-coherent effects. It has been used for exploring potential signatures of unconventional superconductivity and other novel physical phenomena~\cite{Kir11,Gin16,Gos16,Ass19,Pao19,May19a,May19b}. 
For instance, conventional superconductors have been combined in a SQUID geometry with other materials, such as ferromagnets~\cite{Gin16}, topological insulators~\cite{Vel12,Kur15}, or nanowires~\cite{Van06,Ass19}, in order to study non-trivial current-phase relations (CPRs)~\cite{YokNaz14,Yok14,Mar16,Bla19}. 
Additionally, SQUID-based phase-sensitive measurements are an effective tool for investigating more exotic superconductors, such as ruthenates~\cite{Nel04}, LAO/STO interfaces~\cite{Gos16}, or high-Tc cuprates~\cite{Van95,Kir11}.
A SQUID is also the typical framework to study the JJs response individually. For instance, it is possible to create highly asymmetric critical current configurations via voltage gating~\cite{Mon17,Pao19}, or to adjust the direction of the external magnetic drive in order to observe effects dependent on its orientation on the ring plane~\cite{Ass19}. 
Josephson interferometers were also effectively used to study heat currents~\cite{Gia12,Gua18} in phase-dependent caloritronics experiments~\cite{Gia06,ForGia17}, a novel research field dealing with the manipulation of electronic and photonic heat currents in Josephson-based mesoscopic circuits~\cite{Mes06,Gua16,GuaSol18,Gua19,Kam19,Hwa19}.

At the basis of the working principle of a SQUID there is the interference of superconducting wave functions in the two arms of the device, similar to the two slit interference in optics, due to the external out-of-plane magnetic flux piercing the superconducting loop. This leads to a modulation of the critical current of the device with a period of one flux quantum~\cite{Cla04,Gra16}. Furthermore, also the in-plane component of the magnetic field can indirectly play a role, \emph{e.g.}, as long as it affects significantly the properties of the weak links. This is, for instance, the case of a SQUID formed by anomalous JJs~\cite{Buz08}, namely, a $\varphi_0$ SQUID. This particular kind of junctions has a ground state corresponding to a finite phase shift, $\varphi_0$, in the CPR, so that a non-vanishing phase drop can appear also in the absence of a flowing current or, conversely, at a zero phase a current can flow, the so-called anomalous current. In particular, if time reversal and inversion symmetries are broken, junction may develop an anomalous behavior.

Recently, such anomalous phase shift has been experimentally observed in hybrid SQUID configurations fabricated with topological insulator Bi$_2$Se$_3$~\cite{Ass19} and Al/InAs heterostructures~\cite{May19a} and nanowires~\cite{Szo16,Str20}. 
Usually, the combined effect of a non-vanishing ring inductance and anomalous Josephson behavior on an rf-SQUID response is not considered. Instead, we demonstrate that, in response to an in-plane magnetic field, an anomalous phase shift can induce a supercurrent circulating in a SQUID ring with a non-zero inductance.
In this case, the supercurrent contribution to the total magnetic flux can be non-negligible in comparison to the flux due to an applied external magnetic field. Thus, designing a device with a non-zero total inductance of the superconducting ring can be an effective tool to detect and explore the anomalous response of $\varphi_0$ junctions. 

\begin{figure}[t!!]
\centering
\includegraphics[width=\columnwidth]{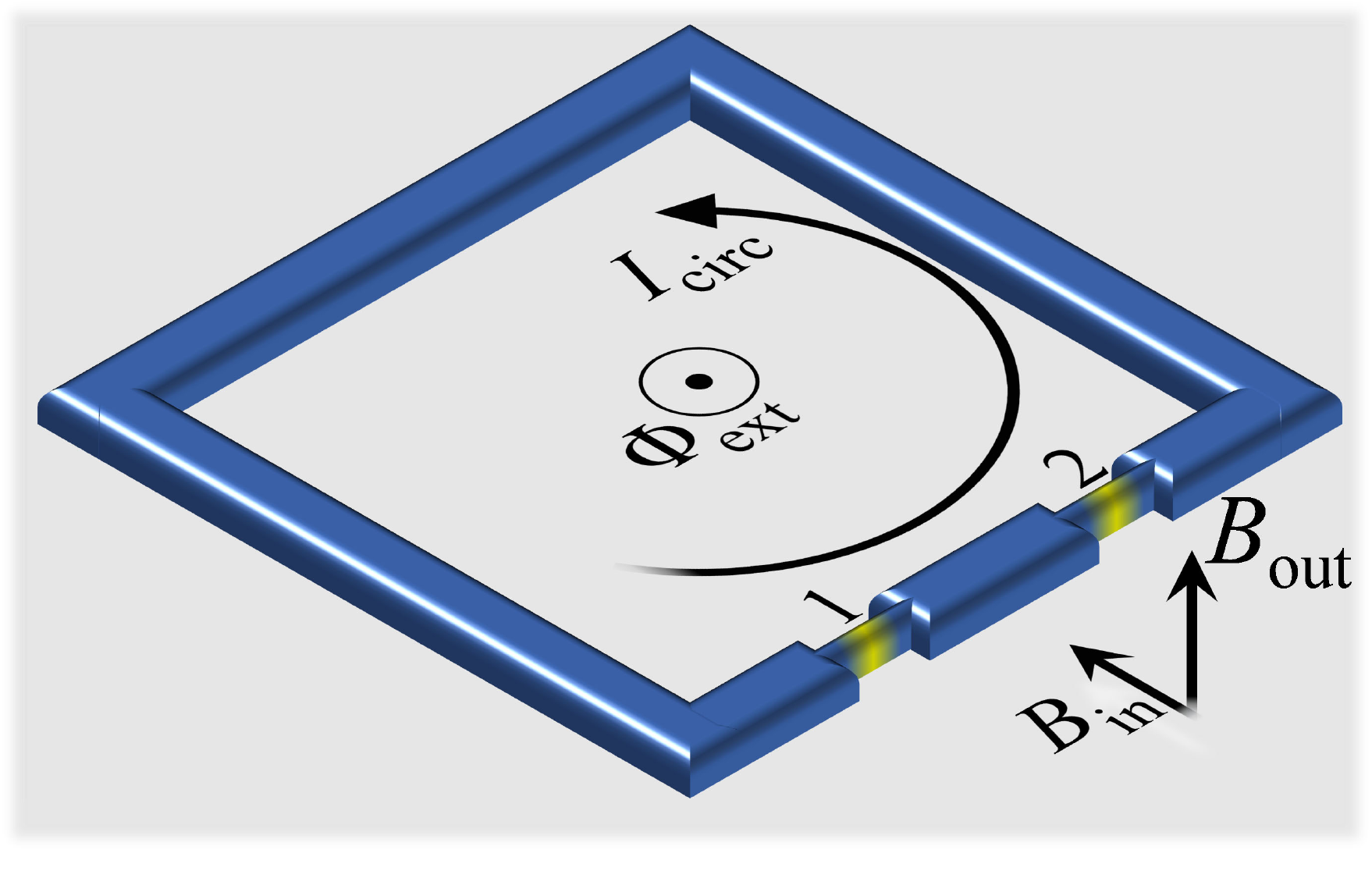}
\caption{Cartoon showing a double junction SQUID affected by both an in-plane and an out-of-plane magnetic field.}
\label{Figure01}
\end{figure}
To be specific, in this work we theoretically study the behavior of a $\varphi_0$ rf-SQUID with a non-negligible ring inductance, when both the out-of-plane and in-plane components of the magnetic field are taken into account. In particular, in the device configurations presented in Fig.~\ref{Figure01}, we show that, by slowly changing the in-plane magnetic field (\emph{i.e.}, in the adiabatic regime), a hysteresis may appear for proper values of the system parameters. This hysteretical mechanism is directly connected to the anomalous phase. We demonstrate also that the magnetic flux through the SQUID ring can be used for further tuning the operating point of the SQUID, in order to avoid (or even to exploit) the hysteretic regions of the device response. Since the magnetic flux at which the system switches to the voltage state depends on $\varphi_0$, it gives a measurement of the anomalous phase. This means that a rf-SQUID can be used to conveniently quantify the anomalous phase shift of the device. Finally, we discuss how this device can allow to investigate the anomalous phase even when the specific dependence of $\varphi_0$ on a control parameter (\emph{i.e.}, the in-plane magnetic field in this work) is not explicitly known. 

The paper is organized as follows. In Sec.~\ref{section1}, the theoretical background used to describe a rf-SQUID with a finite ring inductance formed by $\varphi_0$-junctions is presented. In this section, we explain also how the switching magnetic flux depends on the value of the anomalous phase shift.
In Sec.~\ref{section2}, the response of the SQUID, when only one or both the junctions forming the device are anomalous, is discussed. Here, an effective dependence on the in-plane magnetic field of the anomalous phase shift is taken into account in a prototypical system design. In this section, we also show how this device can be used to study the whole $\varphi_0$ profile. In Sec.~\ref{Conclusions}, conclusions are drawn.

\section{Model and results}
\label{section1}\vskip-0.2cm

In Fig.~\ref{Figure01} we show a SQUID formed by two $\varphi_0$-junctions, under the effects of both an in-plane and an out-of-plane magnetic field. Here, we are dealing with JJs with the ground state corresponding to a finite phase shift, $0<\varphi_0<\pi$, in the CPR, $I_{\varphi}=I_c\sin(\varphi+\varphi_0)$ ($I_c$ is the critical current of the junction). 
The sinusoidal CPR has been shown to describe well some of the experimental works on $\varphi_0$-junctions nowadays available~\cite{Ass19,May19a,Szo16,Str20}. A generalization to a more general non-sinusoidal form is possible, but this requires a numerical treatment different from that one developed in this paper. 
According to the geometry shown in Fig.~\ref{Figure01}, in the case of identical JJs oriented along the same direction (\emph{i.e.}, along the same side of the SQUID) one can assume $\varphi_{0,1}=\varphi_{0,2}$.

Since an rf-SQUID is electrically open, the current circulates only along the ring. It is convenient to define the Josephson phases $\varphi_1$ and $\varphi_2$ so that the Kirchhoff law for the current can be written as 
\begin{equation}\label{Ibias}
I_{c_1}\sin(\varphi_1+\varphi_{0,1})+I_{c_2}\sin(\varphi_2-\varphi_{0,2})=0,
\end{equation}
where $I_{c_i}$ is the critical current of the $i$-th junction. The convention adopted for the Josephson phases and the orientation of the junctions, which are placed on the same side of the ring as shown in Fig.~\ref{Figure01}, imposes that in the previous equation the sign of the anomalous phase of one junction is opposite. 

The degree of asymmetry of the SQUID is accounted by the \emph{asymmetry parameter}, which is defined by the critical currents ratio as
\begin{equation}\label{asymmetryparameter}
\alpha=I_{c_1}/I_{c_2},
\end{equation}
so that Eq.~\eqref{Ibias} can be recast in
$\alpha\sin(\varphi_1+\varphi_{0,1})=-\sin(\varphi_2-\varphi_{0,2})$.

The fluxoid quantization in the SQUID requires that
\begin{equation}\label{fluxoidquantization}
\varphi_1=\varphi_2-2\pi\frac{\Phi}{\Phi_0}+2\pi n,
\end{equation}
where $\Phi _0=h/2e\simeq 2.067\times 10^{-15}\textup{ Wb}$ and $n$ is the amount of flux quanta in the system, which are usually fixed by phase rigidity. Anyway, there are critical situations where the superconducting phase rigidity is temporarily broken, so that a transition $n\to n\pm 1$ may occur, corresponding to the variation of one flux quantum through the superconducting ring. In Eq.~\eqref{fluxoidquantization}, $\Phi$ is the total magnetic flux piercing the SQUID that can be written as follow
\begin{equation}\label{totalflux1}
\Phi=\Phi_{\text{ext}}+LI_{\text{circ}}.
\end{equation}
Here, $\Phi_{\text{ext}}=SB_{\text{out}}$ is the magnetic flux enclosed in the surface $S$ of the SQUID due to the out-of-plane component of the external magnetic field $B_{\text{out}}$, $L$ is the total inductance of the superconducting ring, and $I_{\text{circ}}$ is the current circulating in the loop that reads
\begin{equation}\label{Icirc}
2I_{\text{circ}}=I_{c_1}\sin(\varphi_1+\varphi_{0,1})-I_{c_2}\sin(\varphi_2-\varphi_{0,2}).
\end{equation}
By indicating the total anomalous phase of the SQUID as
\begin{equation}\label{totalanomalousphase}
\varphi_0=\varphi_{0,1}+\varphi_{0,2},
\end{equation}
one can recast the circulating current exactly as
\begin{equation}\label{Icirc2}
\frac{I_{\text{circ}}}{I_{c_1}}=-i_\alpha(\Phi,\varphi_0)\sin\left (2\pi\frac{\Phi}{\Phi_0}-\varphi_0 \right ),
\end{equation}
where
\begin{equation}\label{kalpha}
i_\alpha(\Phi,\varphi_0)=\left [1+\alpha^2+2\alpha\cos\left (2\pi\frac{\Phi}{\Phi_0}-\varphi_0 \right ) \right ]^{-1/2}.
\end{equation}
The reader should note that if the junctions are not oriented as in Fig.~\ref{Figure01}, but are placed in opposite sides of the superconducting ring, the anomalous phases would not simply add as in Eq.~\eqref{totalanomalousphase}, see, \emph{e.g.}, supplemental materials of Ref.~\cite{Str20}.

If one introduces the \emph{screening parameter}
\begin{equation}\label{screeningparameter}
\beta=\frac{2\pi}{\Phi_0}LI_{c_1}, 
\end{equation} 
from Eq.~\eqref{totalflux1} the normalized total flux through the SQUID reads
\begin{equation}\label{totalflux}
\frac{\Phi}{\Phi_0}=\frac{\Phi_{\text{ext}}}{\Phi_0}-\frac{\beta}{2\pi}i_\alpha(\Phi,\varphi_0)\sin\left (2\pi\frac{\Phi}{\Phi_0}-\varphi_0 \right ).
\end{equation} 
In the two limiting cases, that is when $\alpha\to0$ (\emph{i.e.}, a single-junction rf-SQUID) and $\alpha\to1$ (\emph{i.e.}, a perfectly symmetric rf-SQUID), Eq.~\eqref{totalflux} turns into 
\begin{equation}
\Phi=\left\{\begin{matrix}
\Phi_{\text{ext}} -LI_{c_1}\sin(2\pi\Phi/\Phi_0-\varphi_0)\;\;\;\;\text{for}\;\alpha\to0
\\ 
\Phi_{\text{ext}} -LI_{c_1}\sin(\pi\Phi/\Phi_0-\varphi_0/2)\;\;\;\text{for}\;\alpha\to1,
\end{matrix}\right.
\end{equation} 
respectively.

The adiabatic evolution of the SQUID is usually determined by the minimization of its total free energy~\cite{Bo04,Gua17}. The total Josephson energy can be written as
\begin{eqnarray}\nonumber
E_J&=&-\frac{\Phi_0}{2\pi}\left [ I_{c_1}\cos(\varphi_1+\varphi_{0,1})+I_{c_2}\cos(\varphi_2-\varphi_{0,2}) \right ]\\
&=&-\frac{E_{J,0}}{\alpha\; i_\alpha(\Phi,\varphi_0)},\qquad
\end{eqnarray}
where $E_{J,0}=\Phi_0I_{c_1}/(2\pi)$.
Anyway, one needs to take also into account the inductive contribution, due to the screening current flowing into the superconducting ring, so that the free energy reads
\begin{equation}
E=E_J+\frac{LI^2_{\text{circ}}}{2},
\end{equation}
and in the end it can be written as
\begin{equation}\label{freeenergy}
\frac{E}{E_{J,0}}=-\frac{1}{\alpha\; i_\alpha(\Phi,\varphi_0)}+\frac{\beta}{2} i^2_\alpha(\Phi,\varphi_0)\sin\left (2\pi\frac{\Phi}{\Phi_0}-\varphi_0 \right ) ^2.
\end{equation}
\begin{figure}[t!!]
\centering
\includegraphics[width=\columnwidth]{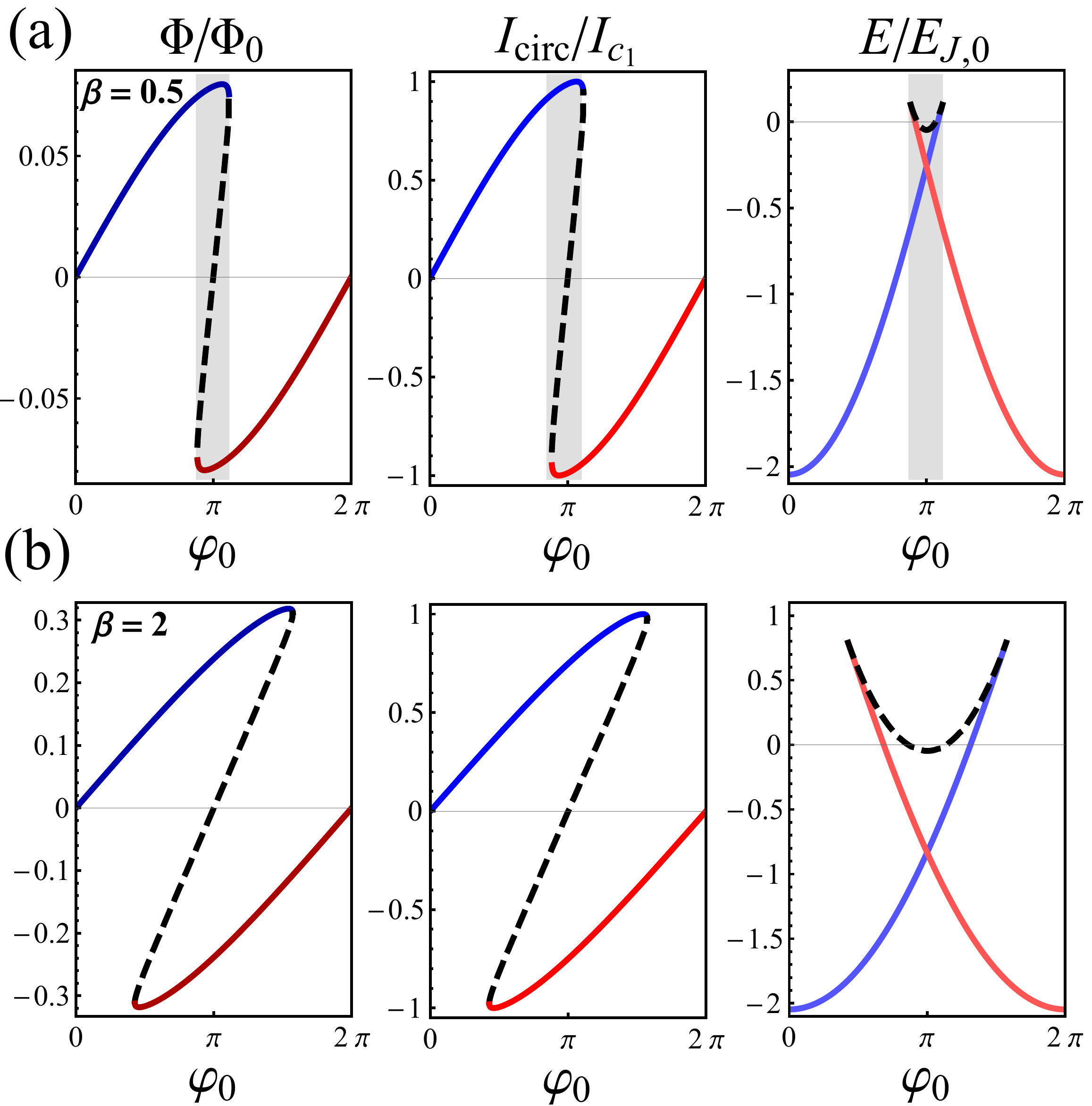}
\caption{Normalized values of total flux, $\Phi/\Phi_0$, circulating current, $I_{\text{circ}}/I_{c_1}$, and free energy, $E/E_{J,0}$ as a function of $\varphi_0$, for $\beta=0.5$ and $\beta=2$ (see top and bottom panels, respectively), at $\alpha=0.95$, in the absence of applied external flux, $\Phi_{\text{ext}}=0$. }
\label{Figure02}
\end{figure}

The typical control parameter adopted to discuss the behavior of a rf-SQUID is the external magnetic flux orthogonal to the superconducting ring, which depends on $B_{\text{out}}$.
The magnetometer output $\Phi\;vs\;\Phi_{\text{ext}}$ of a double junction SQUID with a non-negligible screening parameter $\beta$ is a multiple valued function~\cite{Bar82,Bo04,Gua17}. When the current circulating in the loop equals the lower critical current of one junction, it switches to the normal state, so that the superconducting loop is interrupted and both the phase rigidity and flux quantization conditions cease to hold. Thus, the total flux adjusts to a different value of trapped flux quanta, giving a lower free energy. Interestingly, this process depends on the sweeping direction of the driving external flux, so that in sweeping forth and back $\Phi_{\text{ext}}$ an hysteresis path comes out, with the shape of the hysteretic loop depending on the values of $\alpha$ and $\beta$. Specifically, for $\beta<1-\alpha$ the slope of $\Phi$ is always positive and the $\Phi\;vs\;\Phi_{\text{ext}}$ plot is non-hysteretic. Conversely, for $\beta>1-\alpha$ the slope of $\Phi$ switches from positive to negative, so that $\Phi\;vs\;\Phi_{\text{ext}}$ is multivalued and a hysteretical behavior emerges~\cite{Bo04}. 

We now address the question how the anomalous phase may affect the hysteresis mechanism. The out-of-plane magnetic field, $B_{\text{out}}$ is used to contrast the anomalous phase contribution, giving the possibility to indirectly measure it, as discussed in the following. In Fig.~\ref{Figure02} we show the hysteretic behavior of the total flux, the circulating current, and the free energy as a function of $\varphi_0$. We assume that $\varphi_0$ can be changed by modifying an external control parameter (for instance, we could think to use the in-plane magnetic field, $B_{\text{in}}$). 

We compare the SQUID response assuming two different values of the screening parameter, $\beta=0.5$ and $\beta=2$.  The value of $\beta$ is proportional to the total inductance of the SQUID, but, for a description as close as possible to experimental conditions, we have to consider both the geometric and the kinetic inductances of the superconducting ring. The geometric contribution depends on the design of the loop, so that, for a small superconducting ring, the total inductance should be dominated by the kinetic term, which instead depends on the fraction of condensed Cooper pairs. Indeed the kinetic inductance of a superconducting wire takes the form~\cite{Ann10} $L_K(T)=R_{sq}\frac{l}{w}\frac{\hbar}{\pi\Delta(T)}\frac{1}{\tanh\left [ \frac{\Delta(T)}{(k_BT} \right ]}$, where $R_{sq}$ is the sheet resistance in the non-superconducting state, $l$ and $w$ are the length and the width of the strip, respectively, and $\Delta(T)$ is the temperature-dependent BCS superconducting gap. If we assume a rectangular single-turn loop (with height and width equal to $1\;\mu\text{m}$) of a rectangular wire (the wire thickness and width are equal to $0.05\;\mu\text{m}$) made by Nb (with a typical kinetic inductance per unit length of  $44\;\text{pH}\,\mu\text{m}^{-1}$ at a temperature of $2.5\;\text{K}$~\cite{Ann10}), the geometric and kinetic inductances reads $L_G\simeq2\;\text{pH}$ and $L_K\simeq176\;\text{pH}$, respectively. Thus, if we generically suppose $I_c=1\;\mu\text{A}$ (\emph{i.e.}, a value in line with the critical currents experimentally observed in Refs.~\cite{Ass19,May19a,Str20}), from Eq.~\eqref{screeningparameter} we can reasonably estimate a screening parameter very close to the value $\beta = 0.5$ mainly used in this work.

In the following, we consider almost identical JJs, \emph{i.e.}, we impose $\alpha=0.95$, and we first assume no external flux threading the superconducting ring, that is $\Phi_{\text{ext}}=0$. This means that the hysterical response shown in Fig.~\ref{Figure02} can not be ascribed to the out-of-plane magnetic field. Indeed, the anomalous phase evolution gives rise to a current circulating in the ring, which can produce appreciable effects if the inductance of the SQUID, \emph{i.e.}, the screening parameter, is non-negligible. In fact, if we assume a non-zero $\varphi_0$, a circulating current $I_{\text{circ}}$ flows in the ring, as it is clearly shown in Fig.~\ref{Figure02}(b). The latter generates a non-vanishing total flux through the SQUID, \emph{i.e.}, $\Phi\neq0$. 

\begin{figure}[t!!]
\centering
\includegraphics[width=\columnwidth]{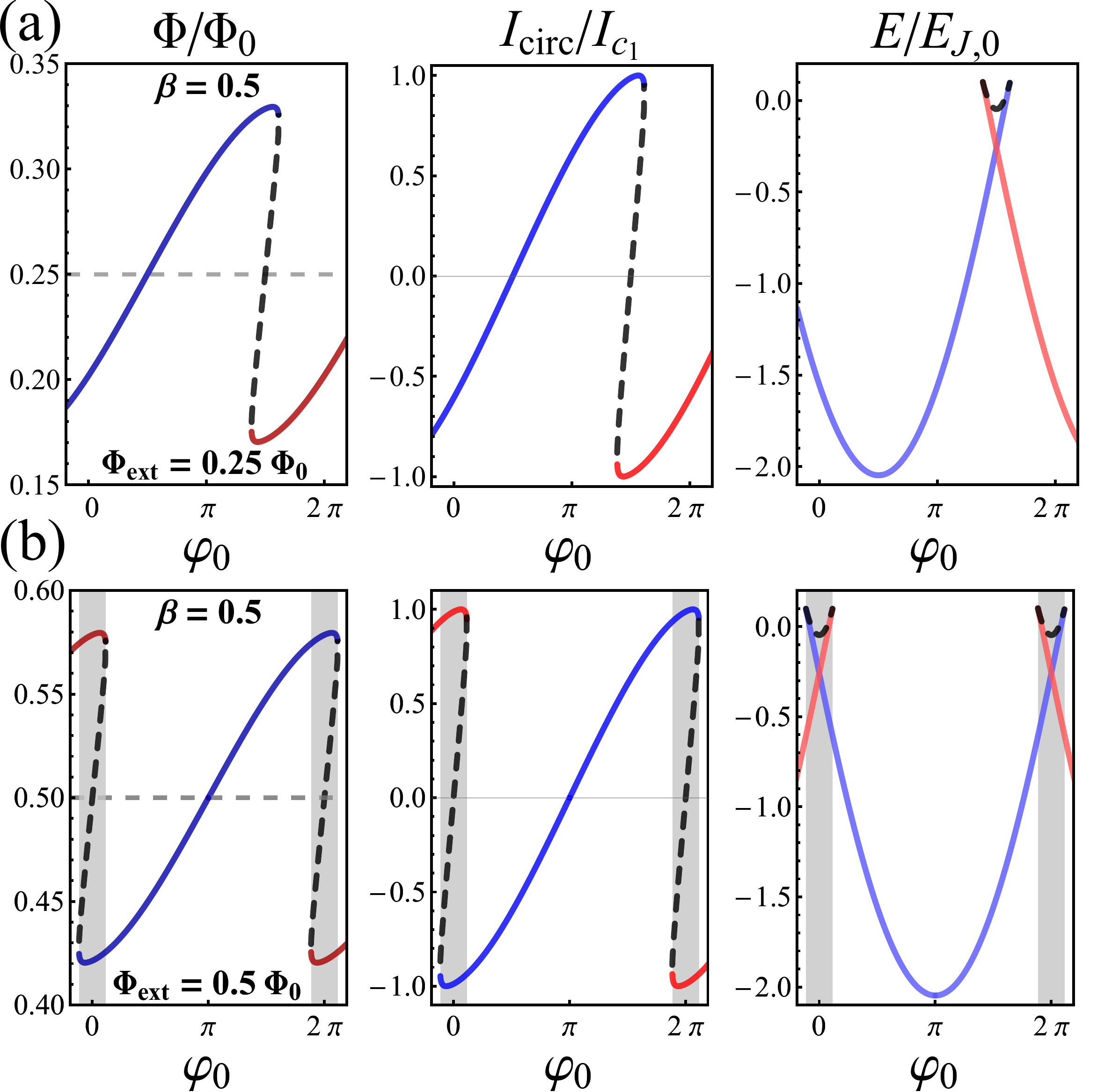}
\caption{Normalized values of total flux, $\Phi/\Phi_0$, circulating current, $I_{\text{circ}}/I_{c_1}$, and free energy, $E/E_{J,0}$ as a function of $\varphi_0$, at $\beta=0.5$ and $\alpha=0.95$, in the presence of an applied external flux with amplitudes $\Phi_{\text{ext}}=0.25\;\Phi_0$ and $\Phi_{\text{ext}}=0.5\;\Phi_0$, see top and bottom panels, respectively. }
\label{Figure03}
\end{figure}

Figures~\ref{Figure02}(c) and (f) show also that the free energy of the system grows with $\varphi_0$, at least as long as $I_{\text{circ}}$ approaches the critical value $I_{c_1}$ (\emph{i.e.}, the lower critical current, since $\alpha<1$). When $I_{\text{circ}}>I_{c_1}$ the device temporarily switches into the voltage state~\cite{Bar82}, a jump to a lower free energy occurs, and the system undergoes a quantum transition $n\to n+1$, \emph{i.e.}, the flux through the SQUID changes by one flux quantum. We mark with a black dashed curves the unstable states of the SQUID, which are not observable during an adiabatic evolution since they have a definitely higher free energy than the stable states of the system. So, sweeping back and forth $\varphi_0$, a hysteretic path is traced out. In fact, after that the increasing of $\varphi_0$ has induced a transition, by reducing further the anomalous phase, the system remains in the $n=n+1$ trapped flux state until the circulating current $\left | I_{\text{circ}} \right |$ reaches again the critical value, so that the free energy suddenly reduces, and the SQUID switches back. Hereafter, dashed curves in the figures serve to indicate unstable states of the system. 

Figure~\ref{Figure02} shows also that the higher is $\beta$, the larger is the hysteretic path traced out by sweeping $\varphi_0$ adiabatically. Moreover, a larger screening parameter gives a larger flux generated by the circulating current, that is, in other words, the maximum value reached by $\Phi$ increases with $\beta$. 
In fact, one can easily compute from Eq.~\eqref{totalflux} the maximum value of the total flux, which reads $ \max_{\varphi_0}\Phi=\frac{\beta}{2\pi}\Phi_0$ in the absence of external field, $\Phi_{\text{ext}}=0$. The anomalous phases $\varphi^+_0$ and $\varphi^-_0$ at which $\Phi$ shows a maximum and a minimum, respectively, can be obtained again from Eq.~\eqref{totalflux} as $\varphi^\pm_0=\pi\pm\left [ \beta-\arccos (\alpha) \right ]$. Thus, the width of the hysteric path can be calculated as
\begin{equation}\label{totalflux2-sol}
\Delta \varphi_0=\varphi^+_0-\varphi^-_0=2\left[ \beta-\arccos (\alpha) \right ].
\end{equation}
Therefore, the higher is the asymmetry of the SQUID (\emph{i.e.}, $\alpha\to0$) and/or the greater is the screening parameter $\beta$, the larger is $\Delta \varphi_0$, that is the wider is the hysteretical path and the more pronounced is the skewing of both the $\Phi$ and $I_{\text{circ}}$ curves, as it is shown in Fig.~\ref{Figure02}. Finally, by increasing $\beta$ the maximum energy approached by the system, and therefore the energy jump when the SQUID undergoes to a $n\to n\pm1$ transition, enhances.

By switching on the external out-of-plane magnetic field, $B_{\text{out}}$, we observe that the overall behavior as a function of the anomalous phase $\varphi_0$ does not change in shape, as it is shown in Fig.~\ref{Figure03}. Here, we assume a SQUID with $\beta=0.5$ under an external flux with intensities $\Phi_{\text{ext}}=0.25\;\Phi_0$ and $\Phi_{\text{ext}}=0.5\;\Phi_0$. In fact, the hysteretical path appears only shifted with respect to the case in the absence of $B_{\text{out}}$. The positions of both the $\Phi$ maximum and minimum are shifted by the same quantity, $2\pi\Phi_{\text{ext}}/\Phi_0$, so that the width $\Delta \varphi_0$ of the hysteretic path is not affected by $\Phi_{\text{ext}}$. 

Usually, in a flux-driven hysteretic SQUID, the total flux $\Phi$ grows less rapidly than $\Phi_{\text{ext}}$, since the flux generated by the screening current opposes $\Phi_{\text{ext}}$~\cite{Gua17}. 
This is also why, in Fig.~\ref{Figure03}(a), at a zero anomalous phase we have a total flux which is well below the external flux value, the latter being indicated by a gray dashed line. In other words, at $\varphi_0=0$ we obtain $\Phi<\Phi_{\text{ext}}$ due to the negative circulating current, see Fig.~\ref{Figure03}(b), that generates a flux which opposes to $\Phi_{\text{ext}}$. 
Since $I_{\text{circ}}<0$ for $\varphi_0=0$, if we assume to adiabatically increase $\varphi_0$ the circulating current increases too, but the condition $I_{\text{circ}}=I_{c_1}$, at which the system jumps at a lower energy state (that is, a more stable state), occurs at a $\varphi_0$ which is higher than that one for the case without external flux. This is why all the curves shown in top panels of Fig.~\ref{Figure03} are shifted towards higher $\varphi_0$ with respect to the cases shown in Fig.~\ref{Figure02}. 

Notably, for $\Phi_{\text{ext}}=0.5\;\Phi_0$ the hysteresis is centered in $\varphi_0=0$ and $2\pi$, see bottom panels of Fig.~\ref{Figure03}. The additional shift in the position of the hysteretic path induced by the external magnetic flux will play a relevant role, especially when the specific magnetic field-dependence of $\varphi_0$ is considered. This situation is discussed in Sec.~\ref{section2}.

\subsection{$\varphi_0$ estimate through a switching flux measurement}
\label{section1b}

The results presented in Figs.~\ref{Figure02} and~\ref{Figure03} demonstrate that an inductive rf-SQUID can exhibit a peculiar response, which is connected to anomalous Josephson effects. In other words, this device offers the concrete opportunity to detect the presence of a phase shift $\varphi_0$ through measurements of magnetic flux, for instance by using another nearby SQUID magnetometer sensor. Alternatively, one can attempt to extract information on the flux behavior by investigating the rf-SQUID in a dispersive configuration, or even through voltage drop measurements in a dc-SQUID setup. In this work, we will not specifically discuss the characteristics of the best detection strategy, which would require a specific design study, but we aim to sketch the best strategy to detect and measure an unknown anomalous phase.

\begin{figure}[t!!]
\centering
\includegraphics[width=\columnwidth]{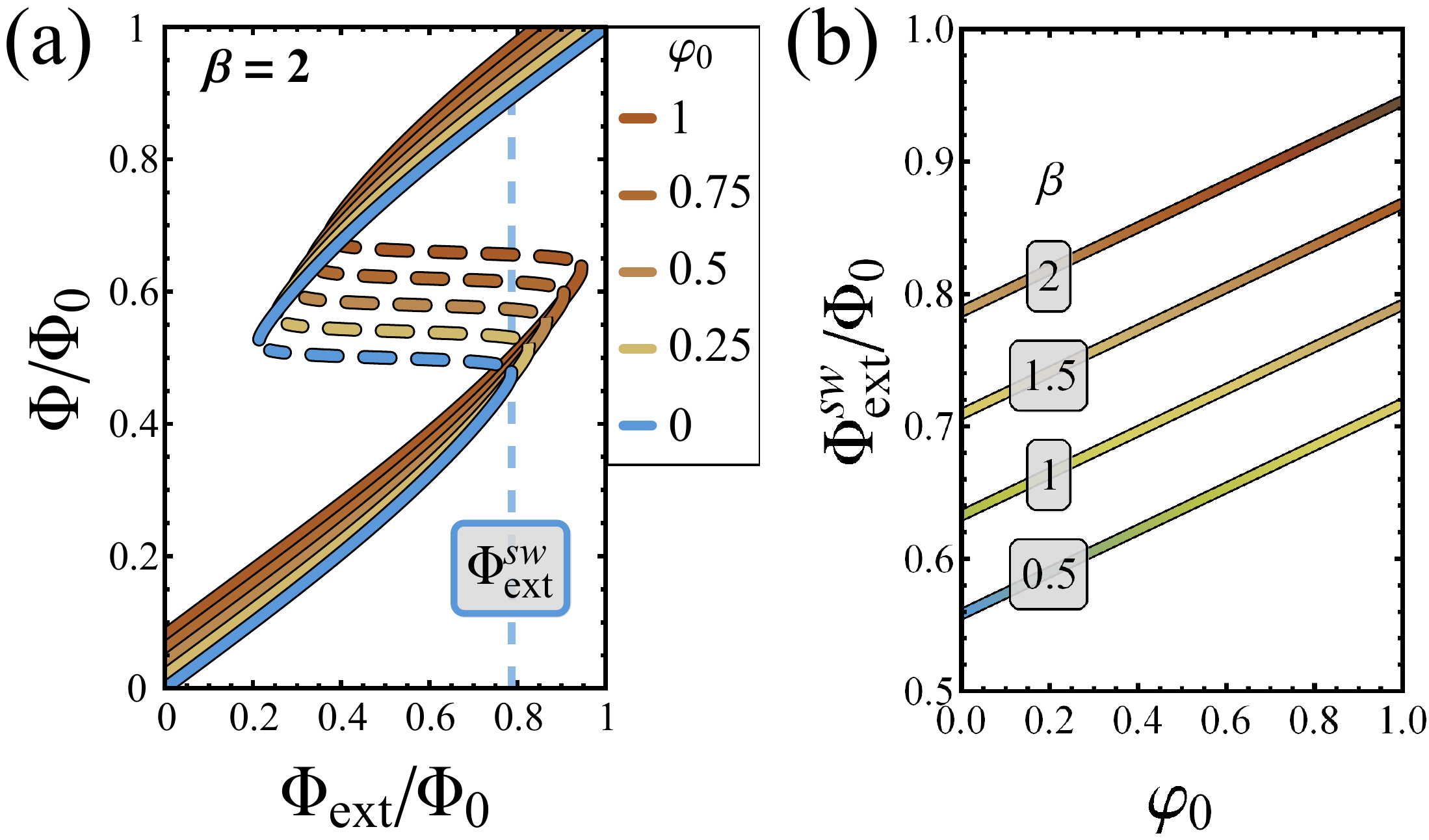}
\caption{(a) Normalized total magnetic flux $\Phi/\Phi_0$ as a function of the normalized external magnetic flux $\Phi_{\text{ext}}/\Phi_0$ at $\beta=2$ and different values of $\varphi_0$. (b) Switching flux $\Phi_{\text{ext}}^{sw}/\Phi_0$ as a function of $\varphi_0$ at different values of $\beta$.}
\label{Figure04}
\end{figure}

Here, we show how the discussed setup can also provide a direct measurement of $\varphi_0$. 
To this aim, in Fig.~\ref{Figure04}(a) we present the behavior of the total normalized flux $\Phi/\Phi_0$ as a function of $\Phi_{\text{ext}}/\Phi_0$, for different values of the phase shift $\varphi_0$, at a fixed $\beta=2$. All curves are multivalued, so that, by adiabatically changing the external magnetic flux, when $\Phi_{\text{ext}}$ approaches the threshold value $\Phi_{\text{ext}}^{sw}$, indicated by a vertical dashed line in Fig.~\ref{Figure04}(a), the system switches to a more stable configuration, with the total flux through the SQUID changing by one flux quantum. When a switch occurs, a detectable voltage drop across the SQUID appears. Thus, Fig.~\ref{Figure04}(a) suggests a way to concretely measure the phase shift, since by changing $\varphi_0$ the switching flux $\Phi_{\text{ext}}^{sw}$ increases too. In particular, the switching flux linearly grows with $\varphi_0$, as it is clearly shown in Fig.~\ref{Figure04}(b). Additionally, in this figure different curves at different $\beta$ are shown. Therefore, by knowing $\alpha$ and $\beta$ (that can be extracted by other standard measurements of the device), an experimental measurement of $\Phi_{\text{ext}}^{sw}$ permits to estimate the anomalous phase shift affecting the device. 

\section{Specific $\varphi_0$ dependence on the external in-plane magnetic field}
\label{section2}\vskip-0.2cm

The results presented in Sec.~\ref{section1} are discussed without specifying how $\varphi_0$ may be eventually generated and modified. The anomalous phase depends on intrinsic properties of the junction, such as the spin-orbit strength and the electron density. In the case of a SOC orthogonal to both the current flowing through the JJs and the in-plane magnetic field (\emph{i.e.}, a SOC colinear with $B_{\text{out}}$ in the geometry depicted in Fig.~\ref{Figure01}), the anomalous phase is a function of $B_{\text{in}}$. We assume the orientation of the in-plane field with respect to the current as shown in Fig.~\ref{Figure01}, so to realize the same assumptions of Ref.~\cite{Ber15}. Thus, according to the discussion in the previous section, a $\varphi_0$-dependent hysteretical response of the SQUID can be induced by applying a magnetic field lying exclusively on the SQUID plane. In this section, we consider a realistic magnetic field-dependence of $\varphi_0$, and we additionally demonstrate that hysteresis with respect to a specific control parameter can emerge also in a device with a quite small screening parameter. We also observe that the range of in-plane magnetic fields within which the system behaves hysteretically is affected by the out-of-plane component of the magnetic drive, thus opening the door to the possibility of reconstructing the magnetic field-dependence of $\varphi_0$ by investigating the hysteretical evolution.

In the diffusive regime, in the presence of a Rashba SOC and a spin-splitting field $h$, the anomalous phase shift has been calculated in Ref.~\cite{Ber15}:
\begin{equation}\label{nonlinearphi0Bin}
\varphi_0=\arctan \left \{ \tanh \left ( \kappa_\alpha L \right )\frac{\sum_\omega \text{Im}\left [ \frac{f_{BCS}^2(\omega)}{\kappa^*\sinh \left ( \kappa^*L \right )} \right ]}{\sum_\omega \text{Re}\left [ \frac{f_{BCS}^2(\omega)}{\kappa^*\sinh \left ( \kappa^*L \right )} \right ]} \right \}.
\end{equation}
Here, $\kappa_\alpha=2\tau\alpha_R^3{m^*}^{2}/\hbar^5$ with $\tau$ and $\alpha_R$ being, respectively, the momentum relaxation time and the Rashba coefficient, $m^*$ is the effective electron mass, $L$ is the junction length, $f_{BCS}(\omega)$ is the BCS bulk anomalous Green’s function in the superconducting leads, $\kappa=\sqrt{\kappa_\omega^2+i\kappa_h^2}$ where $\kappa_\omega^2=2\left | \omega \right |/D$ and $\kappa_h^2=2h/(\hbar D)$ with $D$ and $h=\mu_Bg_sB_{\text{in}}/2$ being the diffusion coefficient and the spin-splitting field, respectively, and the sums extend over the Matsubara frequencies. In the absence of SOC (\emph{i.e.}, $\kappa_\alpha=0$) $\varphi_0$ can be only $0$ or $\pi$. Instead, $\varphi_0$ ranges between $0$ and $2\pi$ for finite values of $\kappa_\alpha$.

For example, in a system with a weak Rashba coupling $\alpha_R$, transparent interfaces, and a negligible spin-relaxation, the anomalous phase shift is simply proportional to the in-plane magnetic field $B_{\text{in}}$. Indeed, by using the same notation of Ref.~\cite{Ass19}, for small $B_{\text{in}}$ Eq.~\eqref{nonlinearphi0Bin} reduces to $\varphi_0=C_{\varphi_0}B_{\text{in}}$, where $C_{\varphi_0}=\frac{\tau {m^*}^{2}\mu_Bg_s(\alpha_RL)^3}{6\hbar^6D}$. Thus, in the case of a $\varphi_0$ linearly dependent on the in-plane magnetic field, the measurement of the switching flux $\Phi_{\text{ext}}^{sw}$ versus $B_{\text{in}}$ discussed in Sec.~\ref{section1b} gives also a direct estimation of $C_{\varphi_0}$, that is, a measurement of the Rashba parameter $\alpha_R$.

Here, we focus on the effects of the in-plane field on the hysteretic behavior of the device, in the case of a fully nonlinear $B_{\text{in}}$-dependence of $\varphi_0$ shown in Eq.~\eqref{nonlinearphi0Bin}. To describe the device response, we need to choose a combination of system parameters, specifically, we impose the values $\kappa_\alpha=0.1\xi_0^{-1}$, $T = 0.1T_c$, and $L = \xi_0$, which are similarly considered in Ref.~\cite{Ber15}. This choice gives a specific $\varphi_0(h)$ profile. We study how this anomalous phase influences the total magnetic flux through the ring, considering two different double-junction SQUID setups, \emph{i.e.}, when only one or both the junctions are anomalous. In both cases, we show that even a rather small screening parameter, \emph{i.e.}, $\beta=0.5$, can produce a non-negligible hysteresis, with and without taking into account an out-of-plane external flux, $\Phi_{\text{ext}}$, piercing the superconducting ring. 

Finally, in Sec.~\ref{Sec3b} we relax the specific choice of parameter values, when we discuss how this device can be used to study the $\varphi_0$ profile, also when the dependence of $\varphi_0$ on a control parameter (for instance, the in-plane magnetic field) is unknown.

\subsection{Single-$\varphi_0$-junction case}
\label{section2a}\vskip-0.2cm

In Fig.~\ref{Figure05} we show the behavior of a double-junction rf-SQUID involving only one $\varphi_0$-junction, so that the total anomalous phase is $\varphi_0=\varphi_{0,1}\in[0-2\pi]$. We investigate the hysteretical response of the device with and without an external magnetic flux piercing the SQUID ring, specifically, we show results for $\Phi_{\text{ext}}=0$ and $\Phi_{\text{ext}}=0.5\;\Phi_0$.

In panels (a) and (c) of Fig.~\ref{Figure05} we present the profile of the anomalous phase $\varphi_0$ as a function of the normalized spin-splitting field $h/\Delta_0$, the latter depending on the in-plane magnetic field (here, $\Delta_0$ is the zero-temperature BCS superconducting gap). In these panels we highlight with horizontal gray shaded bands the range of $\varphi_0$ values within which the SQUID behaves hysteretically. In fact, we have seen that for $\Phi_{\text{ext}}=0$ the system shows hysteresis for $\varphi_0\sim\pi$, see the gray bands in Fig.~\ref{Figure02}(a) and Fig.~\ref{Figure05}(a). Instead, for $\Phi_{\text{ext}}=0.5\;\Phi_0$ the anomalous phases at which hysteresis occurs gather around $0$ and $2\pi$, see the gray bands in Fig.~\ref{Figure03}(d) and Fig.~\ref{Figure05}(c). 
Thus, every time that a given $h$ induces a $\varphi_0$ laying within one of these gray bands, the SQUID can show bistability, which depends on the sweeping direction of the in-plane magnetic field. Hence, in Fig.~\ref{Figure05} we additionally indicate with yellow shaded bands the ranges of $h$ values giving a bistability. The width $\Delta h$ of these yellow bands depends on the slope of $\varphi_0(h)$. Specifically, the lower (higher) the slope of $\varphi_0$, the larger (narrower) is $\Delta h$.
In Figs.~\ref{Figure05}(b) and (d) we show the normalized total flux $\Phi/\Phi_0$ as a function of $h$ at $\Phi_{\text{ext}}=0$ and $\Phi_{\text{ext}}=0.5\;\Phi_0$, respectively. The red and blue curves indicate states of the system with a different amount $n$ of flux quanta in the ring, and a vertical branch indicates a $n\to n\pm1$ transition. For $h$ values within the yellow shaded bands, the bistability of $\Phi$ is quite evident. Interestingly, despite the small value of $\beta$, for $\Phi_{\text{ext}}=0$ the system shows bistability within a quite large range of $h$ values, \emph{i.e.}, for $h/\Delta_0\in[2-15[$, according to the low slope of $\varphi_0$ at $\varphi_0=\pi$. Conversely, for $\Phi_{\text{ext}}=0.5\;\Phi_0$, hysteresis is restricted within small regions at $h\sim0$ and high values of $h$. 

\subsection{Double-$\varphi_0$-junction case}
\label{section2b}\vskip-0.2cm

\begin{figure}[t!!]
\centering
\includegraphics[width=\columnwidth]{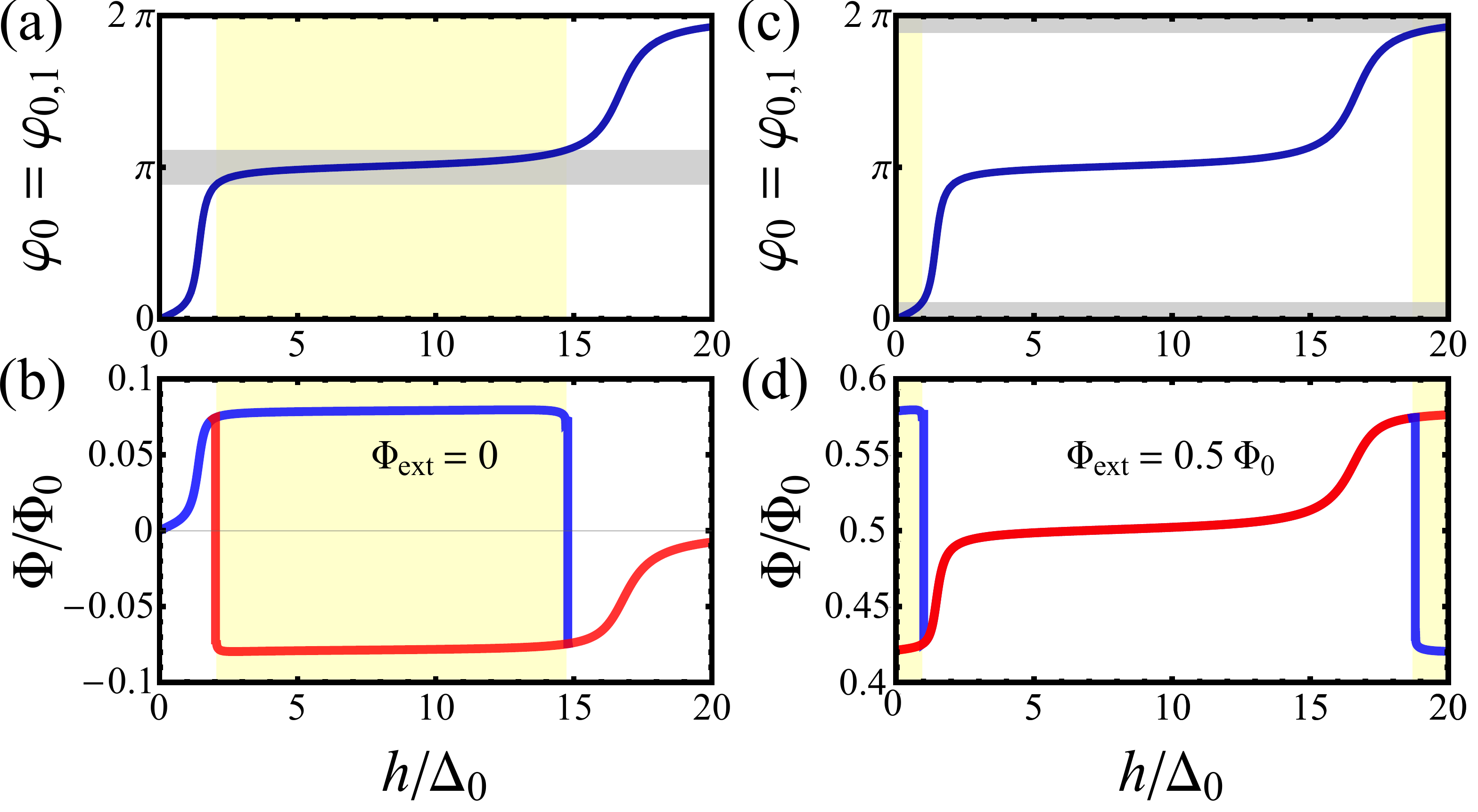}
\caption{Response of an rf-SQUID including only one $\varphi_0$-junction (\emph{i.e.}, $\varphi_{0,2}=0$ and $\varphi_0=\varphi_{0,1}$). (a) and (c), Anomalous phase $\varphi_0$ as a function of the spin-splitting field $h$ (the latter depending on the external in-plane magnetic field), calculated from Eq.~\eqref{nonlinearphi0Bin}. (b) and (d), Total flux through the ring as a function of the spin-splitting field $h$, for $\Phi_{\text{ext}}=0$ and $0.5\;\Phi_0$, respectively. The red and blue curves indicate states of the system with a different amount $n$ of flux quanta in the ring, and a vertical branch indicates a $n\to n\pm1$ transition. The horizontal gray shaded bands in panels (a) and (c) indicate the range of $\varphi_0$ values within which the system behaves hysteretically for $\Phi_{\text{ext}}=0$ and $0.5\;\Phi_0$, respectively [see Figs.~\ref{Figure02}(a) and~\ref{Figure03}(d)]. The yellow shaded bands indicate the ranges of $h$ values giving anomalous phases $\varphi_0$ at which the system behaves hysteretically.}
\label{Figure05}
\end{figure}

\begin{figure}[t!!]
\centering
\includegraphics[width=\columnwidth]{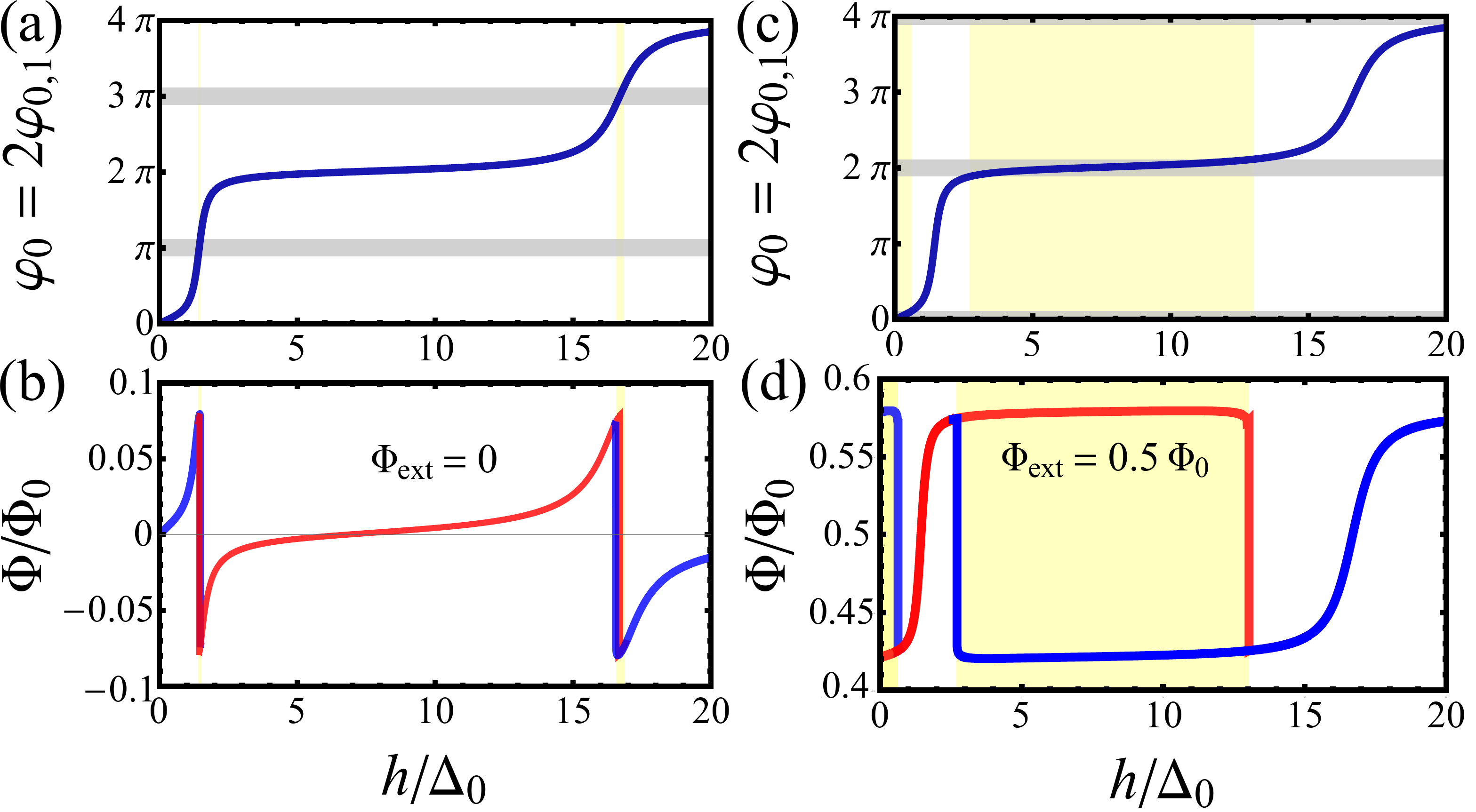}
\caption{Response of an rf-SQUID formed by two identical $\varphi_0$-junctions (\emph{i.e.}, $\varphi_{0,2}=\varphi_{0,1}$ and $\varphi_0=2\varphi_{0,1}$). (a) and (c) Anomalous phase $\varphi_0$ as a function of the spin-splitting field $h$ (the latter depending on the external in-plane magnetic field), calculated from Eq.~\eqref{nonlinearphi0Bin}. (c) and (d), Total flux through the ring as a function of the spin-splitting field $h$, for $\Phi_{\text{ext}}=0$ and $0.5\;\Phi_0$, respectively. The red and blue curves indicate states of the system with a different amount $n$ of flux quanta in the ring, and a vertical branch indicates a $n\to n\pm1$ transition. The horizontal gray shaded bands in panels (a) and (c) indicate the range of $\varphi_0$ values within which the system behaves hysteretically for $\Phi_{\text{ext}}=0$ and $0.5\;\Phi_0$, respectively [see Figs.~\ref{Figure02}(a) and~\ref{Figure03}(d)]. The yellow shaded bands indicate the ranges of $h$ values giving anomalous phases $\varphi_0$ at which the system behaves hysteretically.}
\label{Figure06}
\end{figure}

The scenario presented in the previous section may significantly changes if the SQUID is composed by two identical $\varphi_0$-junctions, see Fig.~\ref{Figure06}. In this case, $\varphi_{0,2}=\varphi_{0,1}$ and the total anomalous phase is $\varphi_0=2\varphi_{0,1}\in[0-4\pi]$, essentially because the junctions operate in series and the anomalous phases add. This strategy can be conveniently adopted to increase the range of value of $\varphi_0$, in the case it is to small to be detected when assuming a single junction. 

In panels (a) and (c) of Fig.~\ref{Figure06} we show the behavior of $\varphi_0(h)$, highlighting again with gray and yellow shaded bands, respectively, the anomalous phases $\varphi_0$ and the corresponding $h$ values at which the system responds hysteretically. 
Since the slope of $\varphi_0$ is quite high at $\varphi_0=\pi$ and $3\pi$, this time for $\Phi_{\text{ext}}=0$ the ranges of $h$ values giving bistability are very narrow, see Fig.~\ref{Figure06}(a). 
Conversely, being the slope of $\varphi_0$ rather small at $\varphi_0=2\pi$, for $\Phi_{\text{ext}}=0.5\;\Phi_0$ the range of $h$ values at which the system is bistable is quite large, see Fig.~\ref{Figure06}(c).

In Figs.~\ref{Figure06}(b) and (d) we display the behavior of the normalized total flux $\Phi/\Phi_0$ as a function of $h$. We note that the hysteresis is negligible for $\Phi_{\text{ext}}=0$, since it comes only out for $h$ values within two narrow ranges. Conversely, for $\Phi_{\text{ext}}=0.5\;\Phi_0$ the SQUID shows bistability in a quite large range of values, \emph{i.e.}, for $h\in[3-13]$, and also in a small extent around $h\sim0$.

Interestingly, for a non-vanishing out-of-plane magnetic field, \emph{i.e.}, $\Phi_{\text{ext}}=0.5\;\Phi_0$ in Figs.~\ref{Figure05}(d) and~\ref{Figure06}(d), both single- and double-$\varphi_0$-junction SQUIDs demonstrate bistability at small in-plane magnetic fields, \emph{i.e.}, around $h\sim0$. In this case, the lower is the slope of $\varphi_0(h)$ at $h=0$, the larger is the hysteretical path. More generally, the in-plane-field-dependence of the hysteretical SQUID response relies strongly on the specific shape of the anomalous phases~\cite{Ber15,Kon15}. The out-of-plane magnetic field can therefore be used as a knob for tuning the best operating point of the device, with the aim to avoid (or, eventually, take advantages of) the bistability, according to the application field for which the device is designed. Furthermore, $\Phi_{\text{ext}}$ can serve also as a tool to investigate how $\varphi_0$ varies as a function of a control parameter, \emph{i.e.}, the in-plane magnetic field in the case discussed in this work. This point is considered in the next section.

\subsection{Measurements of the $\varphi_0(h)$ profile}
\label{Sec3b}\vskip-0.2cm

In Figs.~\ref{Figure05} and~\ref{Figure06}, we highlight with yellow and gray bands, respectively, the width $\Delta h$ of the hysteretical path and the range $\Delta \varphi_0$ of anomalous phases giving bistability. The width of the hysteretical path is a quantity accessible experimentally since it is delimited by two distinct $n\to n\pm1$ dissipative transitions to the voltage state, which give an abrupt change of $\Phi$. The width of gray bands, $\Delta \varphi_0$, depends through Eq.~\eqref{totalflux2-sol} on both $\alpha$ and $\beta$, while its position is a function of the magnetic flux threading the device. For instance, in Fig.~\ref{Figure06} we demonstrated that when $\Phi_{\text{ext}}$ changes from $0$ to $0.5\;\Phi_0$, the gray band moves from $\pi$ to $2\pi$, and the width $\Delta h$ enlarges depending on the inverse of the slope of $\varphi_0(h)$ [\emph{i.e.}, $\Delta h\propto\left (\partial_h\varphi_0  \right )^{-1}$]; in fact, a lower slope gives a larger $\Delta h$. It follows that, by changing $\Phi_{\text{ext}}$ one can shift a gray band and, at the same time, explore how the width $\Delta h$ of the hysteretical path modifies. Thus, by sweeping the magnetic flux threading the SQUID ring, one can outline the profile of the first derivative of $\varphi_0$ with respect to $h$. This means that, in the case of an unknown dependence of the anomalous phase on the in-plane magnetic field, one can ``reconstruct'' the profile of its first derivative, at least if the range of anomalous phases giving hysteresis is small enough. The width $\Delta \varphi_0$ of the gray band is imposed by structural properties of the SQUID, since it directly depends on both $\alpha$ and $\beta$ (the latter being determined by the critical currents and the total inductance of the superconducting ring). In other words, these system parameters set the resolving power of the detection method.

Notably, this inspection scheme of anomalous Josephson effect has a broad versatility since it does not depend on the specific control parameter used to drive the anomalous phase. In fact, in the place of the in-plane magnetic field, one can alternatively use gate voltages applied to the weak links, in order to tune the Rashba coefficient $\alpha_R$ and, consequently, the anomalous phase $\varphi_0$.

To sum up the potential applications of our rf-SQUID-based technique for investigating anomalous Josephson effects, the setup described in our work permits not only to establish the value of the anomalous phase through measurements of the switching magnetic flux, as discussed in Sec.~\ref{section1b}, but it can also serve to study the anomalous Josephson effect in the whole sense when the specific dependence of $\varphi_0$ on a control parameter (\emph{e.g.}, an in-plane magnetic field or a voltage gating) is not known \emph{a priori}. 

\section{Conclusions}
\label{Conclusions}\vskip-0.2cm

In this paper, we study the hysteretical response of an inductive $\varphi_0$ rf-SQUID, namely, a superconducting ring with a non-negligible inductance interspersed by $\varphi_0$-junctions, when both the in-plane and the out-of-plane magnetic field components are taken into account.

The emergence of this hysteretic mechanism dependent on the in-plane magnetic field, and the feasibility of controlling the SQUID response via the external magnetic flux piercing the device, opens the door to alternative ways for investigating anomalous phases by sensing the total magnetic flux, or by measuring the value of the external magnetic flux at which the SQUID switches to the voltage state.

Specifically, in the setup shown in Fig.~\ref{Figure01}, we analyze the total flux $\Phi$ through the device, the current $I_{\text{circ}}$ circulating in the superconducting ring, and the free energy of the system as a function of the total anomalous phase shift $\varphi_0$. By changing the values of the screening parameter $\beta$, which is proportional to the product of the ring inductance and the Josephson critical currents, we observe that a larger $\beta$ gives both a larger range of bistability and a more pronounced skewing of both the $\Phi$ and $I_{\text{circ}}$ curves. Additionally, we show that the switching flux, that is, the external magnetic flux value at which the system switches to the voltage state, is linearly proportional to the anomalous phase shift of the device.

With the aim to explore the range of values within which the system shows bistability, we assume a specific dependence of the anomalous phase shift on the in-plane magnetic field. Besides, we study two different system setups, \emph{i.e.}, when only one or both the junctions forming the SQUID are anomalous. We observe that the response of the system to the in-plane magnetic field, and therefore also the hysteresis phenomenon, strongly depends on the device configuration. Thus, we discuss how the out-of-plane magnetic field can be employed to set the working regime of the device, in order to eventually avoid hysteresis. 
Finally, we proposed a detection scheme based on a $\varphi_0$ rf-SQUID that allows to measure the anomalous phase and also to determine its functional dependence on an external control parameter, such as the in-plane magnetic field or a gate voltage applied to the junctions.

Finally, we remark that the phenomenology presented in the manuscript is not strictly dependent on the choice of a sinusoidal CPR. The main goal of our work is not to discuss the effect of the specific shape of the CPR, but rather the consequence of an anomalous phase on the magnetic SQUID response in the case of a non-vanishing ring inductance. This result is connected more with the fact that at zero phase bias, $\varphi = 0$, a finite (anomalous) current may exist, than on the specific form of the CPR, which is assumed sinusoidal for simplicity. Further non-sinusoidal higher order corrections can become important to predict the specific values of the magnetic flux, as much as the proper strategy to measure an anomalous Josephson effect, but do not spoil the sensitivity of the proposed setup to the anomalous current detection. 
However, we are aware that in the case of a non-sinusoidal skewed CPR, such us in systems with a very high transparency~\cite{Kon15,May19a}, the study can be eventually performed using purely numerical methods, for instance, through a time-dependent resistively and capacitively shunted junction (RCSJ) model~\cite{GuaVal15,GuaVal17,GuaVal19,Gua20} or along the line of the work done by Podd \emph{et al.} for micro-SQUIDs~\cite{Pod07}. We reserve such analysis for a future research.

\begin{acknowledgments}
F.S.B., C.S.F., and C.G.  acknowledge funding by the Spanish Ministerio de Ciencia, Innovaci\'on y Universidades (MICINN) (Project No. FIS2017-82804-P). FSB aknowledge  partial support  by Grupos Consolidados UPV/EHU del Gobierno Vasco (Grant No. IT1249-19), and by EU's Horizon 2020 research and innovation program under Grant Agreement No. 800923 (SUPERTED). 
The work of E.S. was supported by a Marie Curie Individual Fellowship (MSCA-IFEF-ST No.660532-SuperMag). 
E.S., N.L, and F.G acknowledge partial financial support from the European Union’s Seventh Framework Programme (FP7/2007-2013)/ERC Grant No. 615187- COMANCHE. 
E.S., A.I., N.L, F.S.B., and F.G were partially supported by EU’s Horizon 2020 research and innovation program under Grant Agreement No. 800923 (SUPERTED). 
A.B. acknowledges the CNR-CONICET cooperation program Energy conversion in quantum nanoscale hybrid devices, the SNS-WIS joint lab QUANTRA, funded by the Italian Ministry of Foreign Affairs and International Cooperation and the Royal Society through the International Exchanges between the UK and Italy (Grant No. IES R3 170054 and IEC R2 192166). 
\end{acknowledgments}


%

\end{document}